\documentclass[pdftex]{JINST}
\pdfoutput=1

\title{Design of the MiniCLEAN dark matter search veto detector subsystem}
\author{Robert~Abruzzio$^a$, Benjamin~Buck$^a$\thanks{Corresponding author.}, Stephen~Jaditz$^a$, James~Kelsey$^a$, Jocelyn~Monroe$^{b}$, Kimberly~Palladino$^{c}$\\
\llap{$^a$}Department of Physics, Massachusetts Institute of Technology,\\
	Cambridge, MA, USA\\
	E-mail: \email{bbuck@mit.edu}\\
\llap{$^b$}Department of Physics, Royal Holloway University of London,\\
	Egham, Surrey, UK \\
\llap{$^c$}SNOLAB, \\ 
        Lively, Ontario, CA} 

\abstract{This paper describes the design of the active muon veto subsystem for
the MiniCLEAN dark matter direct detection experiment at SNOLAB in Sudbury,
Ontario, Canada. The water-filled veto is instrumented with 48 PMTs which are
read out by front end electronics to time multiplex 48 photomultiplier channels
into 6 digitizer channels and provide an instantaneous hit sum across the
subsystem (N-Hit) for the veto trigger. We describe the primary system components:
the PMTs, the support structure, the front-end electronics, and the data acquisition system.}

\keywords{Dark matter detectors, Photon detectors for UV, visible and IR photons (vacuum), Front-end electronics for detector readout}

\begin{document}

\section{Introduction}
\label{Introduction}
The MiniCLEAN experiment is searching for Weakly Interacting Massive Particle
(WIMP) dark matter using a liquid argon target. The MiniCLEAN detector
physics goals and design are described in~\cite{ref:miniclean_physics}. The
detector is located 6800 feet underground in the Cube Hall at SNOLAB in
Sudbury, Ontario, Canada. The detector consists of a spherical stainless
steel inner vessel containing liquid argon or neon surrounded by 92 cryogenic
photomultiplier tubes (PMTs) inside an outer vacuum vessel. The outer vessel
is located within a water tank, shown in Figure~\ref{fig:veto_geom}, which
provides additional background shielding as well as a veto function for
external particles incident on the argon detector volume. 

The veto water shields the liquid argon target from low energy neutrons and gamma
rays produced by radioactivity in the surrounding cavern rock and tags cosmic muons
which penetrate the rock overburden to the detector depth. These muons can
create high energy neutrons, which may mimic the WIMP signal.
From a simulation of the cosmogenic muon flux, using the Sudbury Neutrino Observatory (SNO) measurement of
3.31$\pm$0.01(stat.)$\pm$0.09(sys.)$\times$10$^{-10}$ $\mu$/cm$^2$/s
~\cite{ref:sno_muon_flux}, and the energy and angular distribution
parameterization from~\cite{ref:mei_and_hime}, the expected total muon flux
through the MiniCLEAN veto tank water is 9.8 muons/day. 48 PMTs are used to
detect Cherenkov light produced by muons transiting the water, and trigger a
veto signal. The veto was designed to tag 99.9\% of cosmic muons, and GEANT4
~\cite{ref:geant4} simulations were done to determine the number and placement
of PMTs inside the water volume. 

The MiniCLEAN Muon Veto Subsystem consists of:
\begin{itemize}
\item 48 PMTs mounted on 12 strings of 4 PMTs around the inner diameter of the MiniCLEAN water shield tank;
\item high voltage power supply for the PMTs; and,
\item veto PMT signal electronics, which include custom amplifier discriminator boards, custom summer boards, and one CAEN V1720 digitizer board for data acquisition. 
\end{itemize}
The custom electronics bias the PMTs, time multiplex 8 PMT channels into a
single digitizer channel, and generate a signal proportional to the number
of PMTs above threshold at any moment which is used as the veto trigger to
the DAQ. An overview of the system is shown in
Figure~\ref{fig:block_diagram}.

This paper describes the mechanical and electrical design of the veto
subsystem. The mechanical system is described in
Section~\ref{sec:subsystem_design}, and the electronics are
described in Section~\ref{sec:electronics_design}.

\section{Mechanical design of the MiniCLEAN muon veto subsystem}
\label{sec:subsystem_design}
\begin{figure}[ht]
\begin{center}
\includegraphics[width=3.5in]{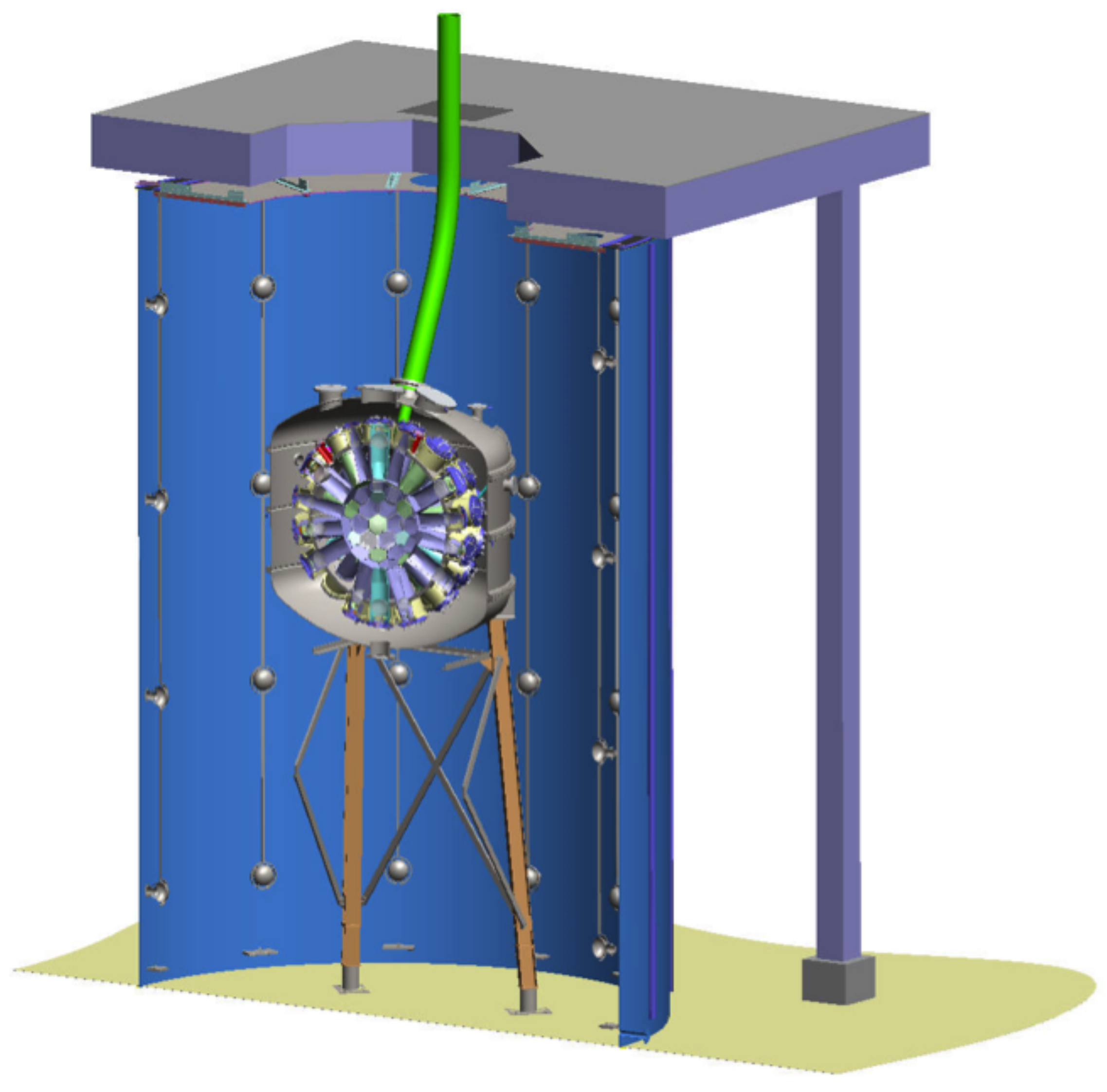}
\caption{Model of MiniCLEAN veto showing the shield tank, inner liquid argon detector, and veto PMTs.
\label{fig:veto_geom}}
\end{center}
\end{figure}

The water tank for the MiniCLEAN veto is a bottomless silo, with
radius 2.8~m and height 7.9~m. These dimensions provide 1.5~m or
greater water thickness between the cavern air and outer vessel. The
veto interior surface is covered by a highly-reflective waterproof liner, with four
holes in the bottom through which the supports for the main detector
pass into the cavern floor. Four 8-inch Hamamatsu R1408 PMTs from SNO are
attached to each of 12 equally spaced poles which hang in the water
along the inside wall of the tank. GEANT4~\cite{ref:geant4} simulations
of the veto performance were used to optimize the mechanical
design. All PMTs point inward, normal to the tank wall. The heights
of the PMTs are the midpoints of four equal lengths extending along
the entire height of the tank. The veto configuration is shown in
Figure~\ref{fig:veto_geom}. 

The simulation of this geometry and the Hamamatsu R1408 PMT response (described
in Section~\ref{sec:pmts} results in an average of approximately 25
photoelectrons (p.e.) detected per PMT per muon in the water. The
average time spread between first and last detected p.e. of the signal
on a given PMT is $\sim$15~ns. The electronics described in
Section~\ref{sec:Amp-Disc} are designed for a 0.25~p.e. threshold (2.5~mV)
per PMT; requiring a coincidence of at least 2 PMTs above this
threshold (N-Hit$>$2) results in an efficiency of 0.999 giving 3
un-vetoed muons per year. For comparison, the simulated efficiency
drops to 0.99 if the N-Hit threshold is 1 p.e. and the veto trigger
requirement is N-Hit$>$8. The un-vetoed muons traverse less than 5~cm
of water, clipping the corners of the veto tank volume. From a
simulation of 75 years of cosmogenic muon backgrounds, the estimated
number of cosmogenic background induced scatters in the liquid argon
target with 20-100~keVee is $<$0.1/year.

\subsection{PMTs, connectors, cables}
\label{sec:pmts}
\begin{figure}[ht]
\begin{center}
\includegraphics[width=4.5in]{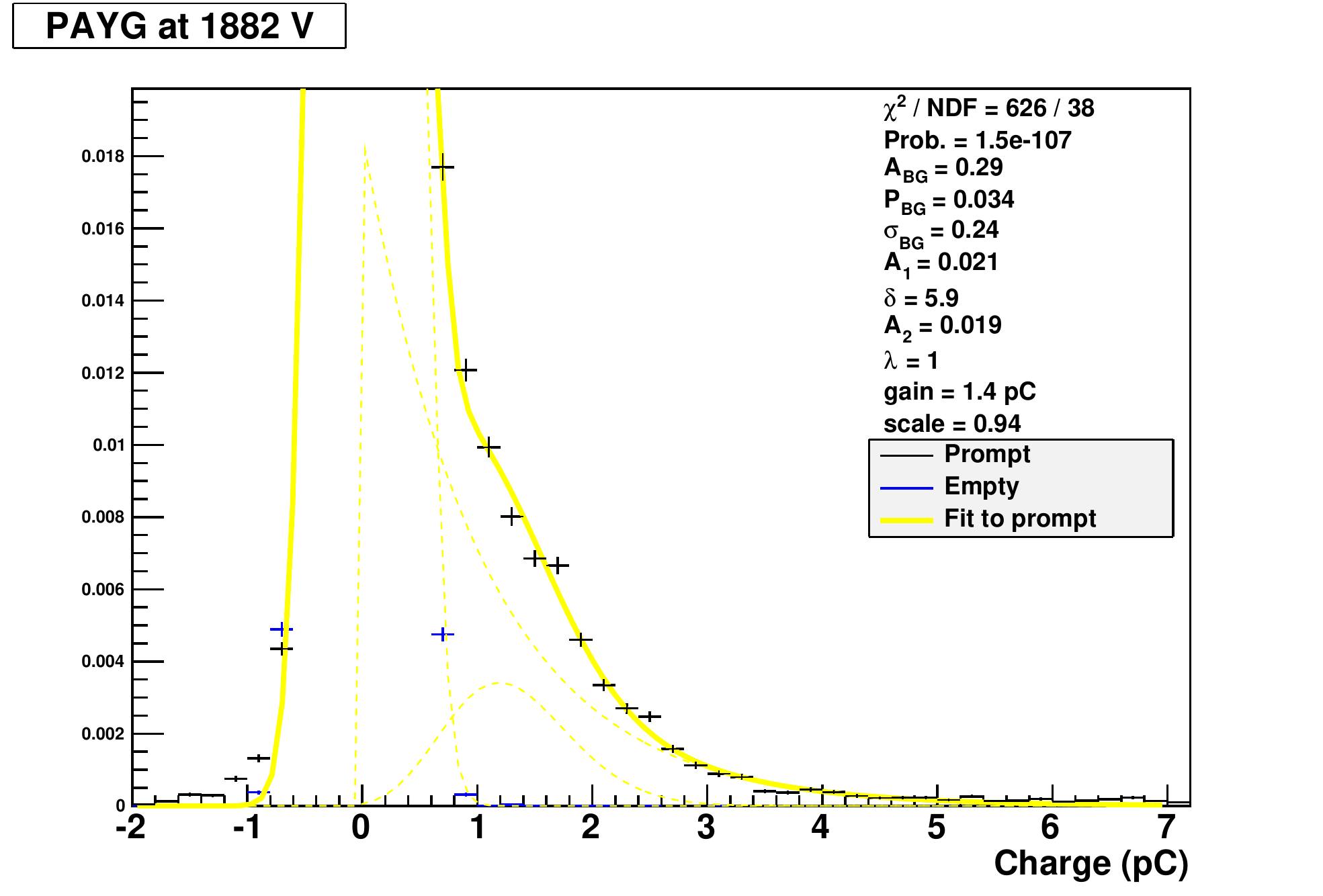}
\caption{Single photoelectron charge spectrum for PMT ID PAYG at operating voltage. The data with a pulsed LED are shown in black, with statistical errors, noise data are shown in blue, and the fit model is shown in the yellow line.
\label{fig:pmt_spe_q}}
\end{center}
\end{figure}

66 Hamamatsu R1408 PMTs were made available from SNO for use in the MiniCLEAN
veto. The gain and dark rates of these PMTs were measured as a
function of bias voltage in a PMT test stand at Los Alamos National
Laboratory. We used a CAEN V1720 digitizer with the MiniCLEAN DAQ
software, DCDAQ, for data acquisition \cite{ref:gastler_thesis}. PMTs were conditioned at 2100~V
for 12 hours. Next, a single photoelectron gain measurement as a
function of bias voltage was made using a blue LED flashed at low
intensity, such that PMT occupancy was several percent. Final gain and
dark rate measurements were made in a similar test stand at MIT, and
48 PMTs plus 5 spares were selected to use in the veto. The selection
criterion was to chosen to minimize dark rate. We required the rate of pulses with
$>$3~mV (approximately 0.3~p.e.) to be less than 3000~Hz in the
absence of a light source. Operating voltages were chosen to gain
match the PMTs for an anode gain of 1$\times$10$^7$; the range of
operating voltages is 1800-2300~V. A typical single p.e. pulse charge
is $\sim$1~pC, with pulse amplitude of $\sim$10~mV and pulse full width of
$\sim$10~ns. These PMTs were also qualified in tests done many years ago for
the Sudbury Neutrino Observatory (SNO), and the optimum operating voltage was
found to be consistent with these earlier measurements.
An example PMT single p.e. charge spectrum at its operating voltage is
shown in Figure~\ref{fig:pmt_spe_q}.

\begin{figure}[ht]
\begin{center}
\includegraphics[width=2.75in,page=13]{graphics/snoConnectorDrawings.pdf}
\includegraphics[width=2.75in]{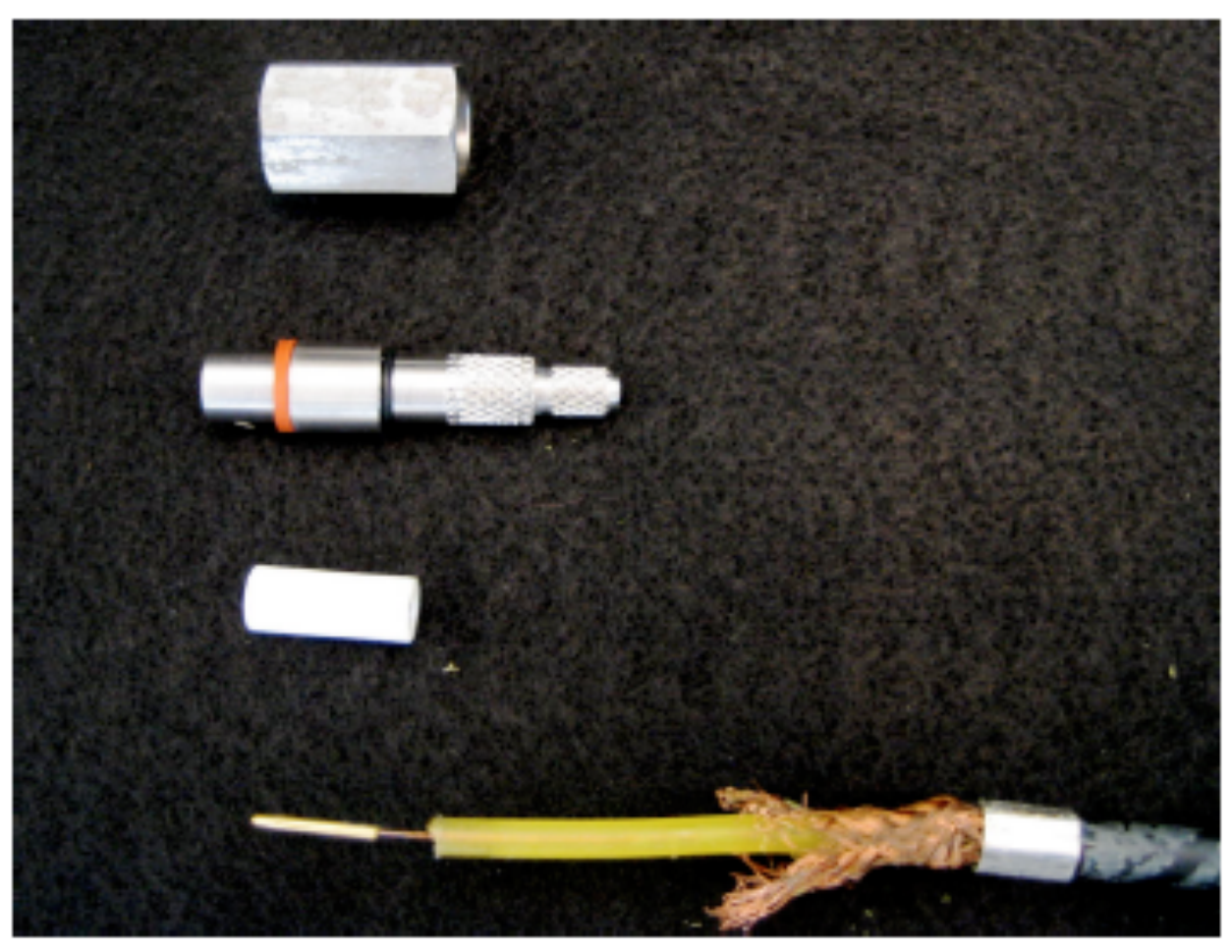}
\caption{Left: TRIUMF connector assembly drawing. Items 1,2,4,6,7,10, and 11 were built or purchased to mate RG59 cable to the existing R1408 PMT jacks. Right: As-built TRIUMF connector for mating cable to the R1408 jack. From top to bottom is shown the coupling nut, the plug body into which the cable is inserted, the teflon insulator which is also inserted into the plug body, and the pin with a stripped cable.
\label{fig:connectordrawing}}
\end{center}
\end{figure}

The SNO PMT bases are fitted with custom, waterproof TNC-type jack
connectors which mate 75 Ohm RG59 cable to the PMTs. These custom
connectors have been built by the MIT machine shop from the SNO-TRIUMF
design. An assembly drawing of the connector is shown in
Figure~\ref{fig:connectordrawing} along with a picture of one of those
built at MIT. The MIT-built SNO
connectors were prototyped, tested at voltage in air, and tested at
voltage in water before the full connector complement was machined.
After cable assembly, all of the machined connectors were tested in
a test stand at RHUL in water at voltage with PMTs attached. All of the
PMTs were tested in water at voltage with the MIT-built connector
sample. Several failures of cable connectors were observed in dry
tests. These experiences were used to modify the cable assembly
procedure after which no cable connectors or PMTs were found to fail
these wet tests.

The connection to the PMTs uses a custom cable, Belden part number YR29304 Rev 2. The cable is a
RG-59 coaxial cable and is the same cable used by SuperK and SNO.
The cable is flooded, has a waterproof jacket and is rated to 2.3~kV.

The SNO PMTs come with waterproof base enclosures, consisting of a
plastic cylinder filled with gel. All of the PMTs were immersed in water and operated at voltage for approximately 4 hours in a test stand at
MIT to look for base waterproofing or connector failures. However,
this test did not simulate the water pressure at the bottom of the
veto tank. To address this question, a pressure vessel was assembled
at MIT Bates Research and Engineering Center to test the PMT bulbs, base enclosures, connectors, and
cables for waterproofness and structural integrity at pressure
equivalent to that at the bottom of the veto. Testing
all of the PMTs at atmospheric pressure was successful and a random
subset of 10\% of the PMTs tested at 4 atm for several weeks (a factor of 2
above the maximum pressure in the MiniCLEAN veto tank) found no
failures. Therefore we concluded that further encapsulation of the
SNO waterproof base enclosure was not necessary.

\subsection{Mechanical hardware}
\begin{figure}[ht]
\begin{center}
\includegraphics[width=2.in]{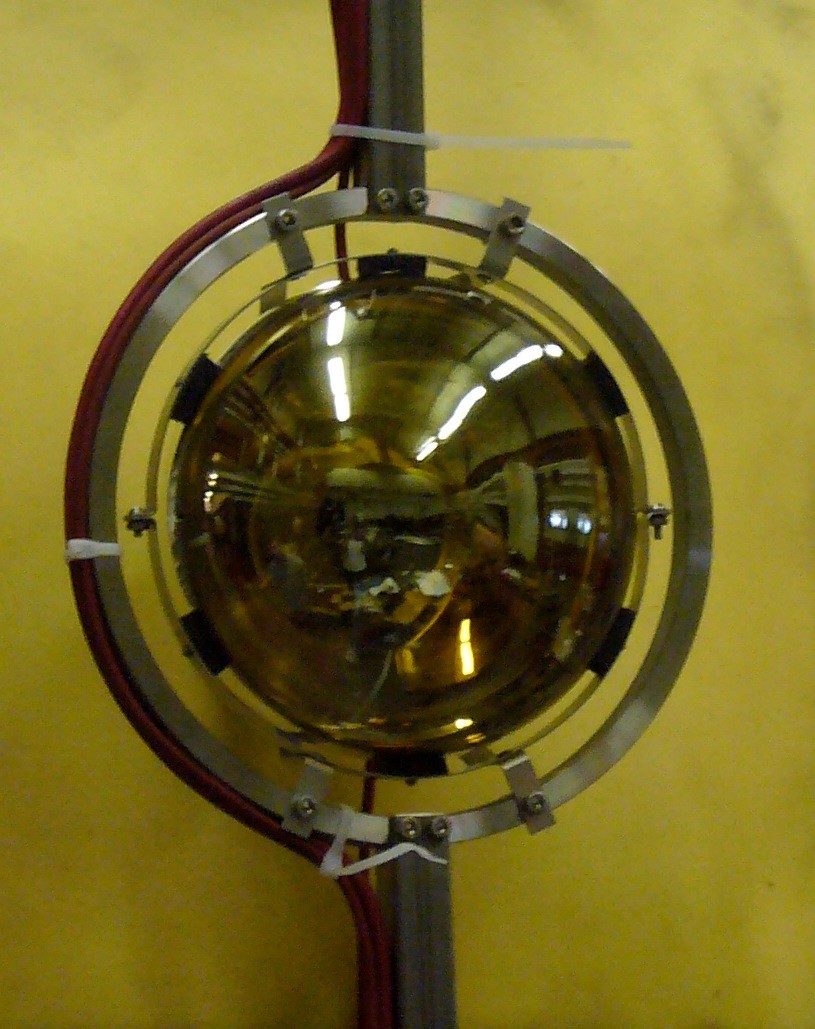}
\includegraphics[width=1.9in]{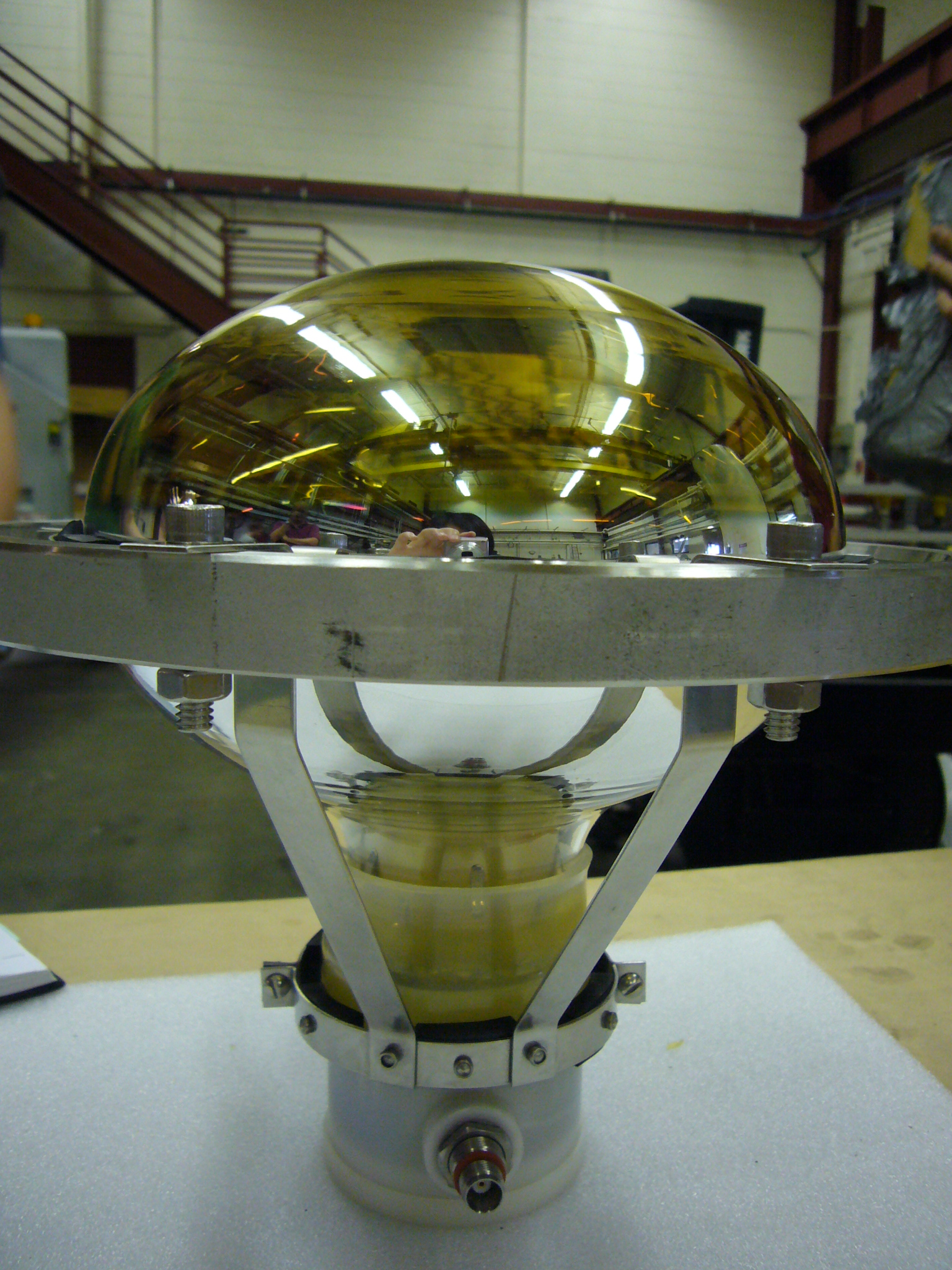}
\includegraphics[width=1.9in]{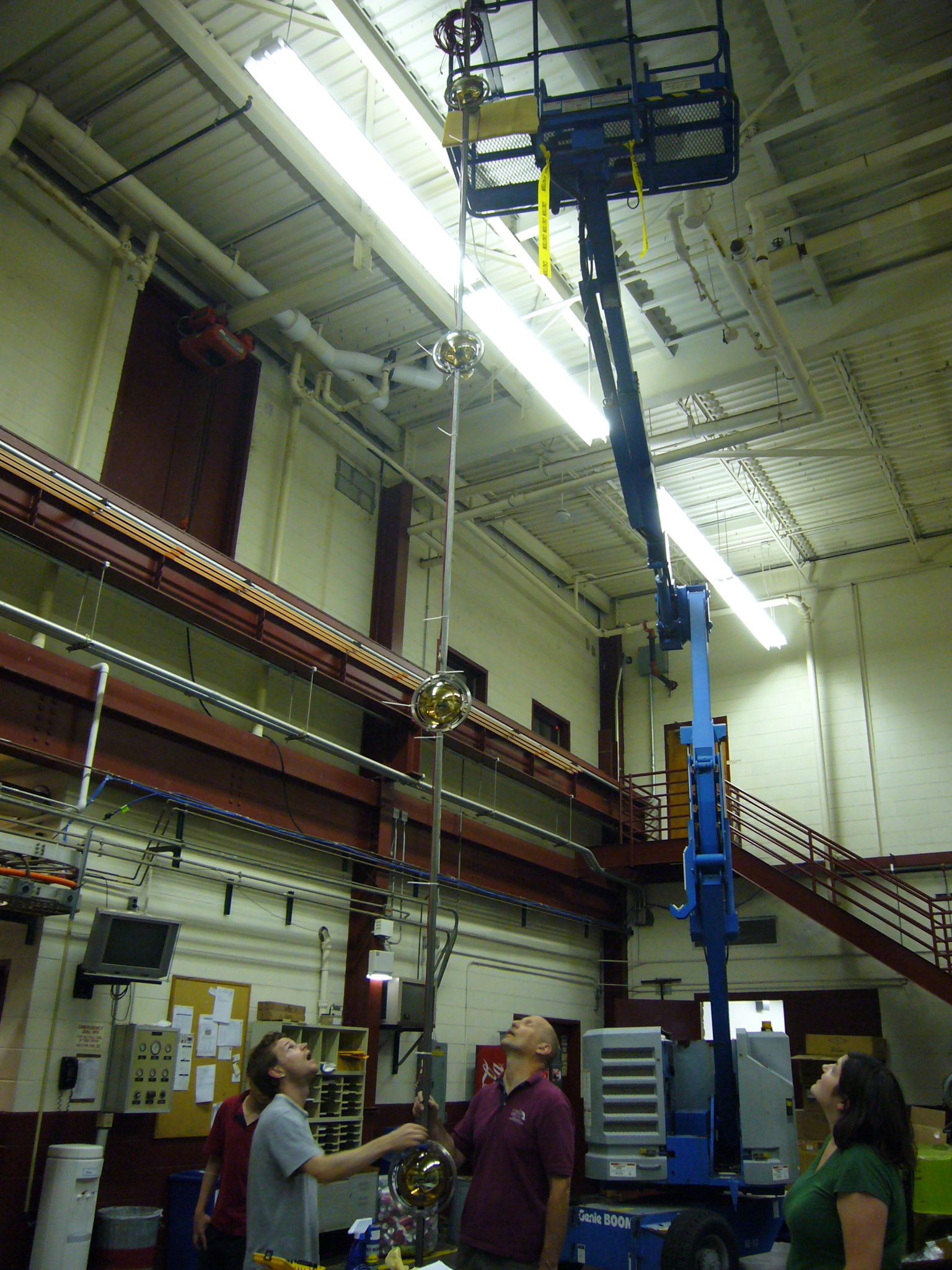}
\caption{Veto PMT shown in mounting hardware (left, center), and assembled into a 4-PMT string (right).
\label{fig:vetopmtmountpic}}
\end{center}
\end{figure}

The PMT support structures attach the PMTs to poles or
``strings'' which attach to the veto tank lid. These have been built at
MIT Bates Research and Engineering Center. 
The 316L stainless steel mount and string are shown in
Figure~\ref{fig:vetopmtmountpic}. The
Buna-N rubber pads contacting the base enclosure and bulb accommodate
the variations in dimensions across the set of PMTs. The base
enclosures' and bulbs' diameters were measured to vary between 3.00
and 3.34 inches and between 7.91 and 8.02 inches. The rubber mounting
pads were all fit by hand to each PMT at Bates before shipment to
SNOLAB and tested with a load of twice the buoyant force (maximum
8~lbs/PMT) to try to dislodge the PMT from the mount. This procedure
will be repeated during assembly to verify the mechanical stability of
each veto PMT in its mount. These tests were done with both wet and dry
rubber pads. A complete assembly test of one string, consisting of
four PMTs and associated hardware, dry, was done Bates (shown in Figure~\ref{fig:vetopmtmountpic}, right).

\section{Electronics design of the MiniCLEAN veto subsystem}
\label{sec:electronics_design}
\begin{figure}[ht]
\begin{center}
\includegraphics[width=5in, keepaspectratio=true, trim=1.25in 5.75in 0.5in 2in, clip=true]{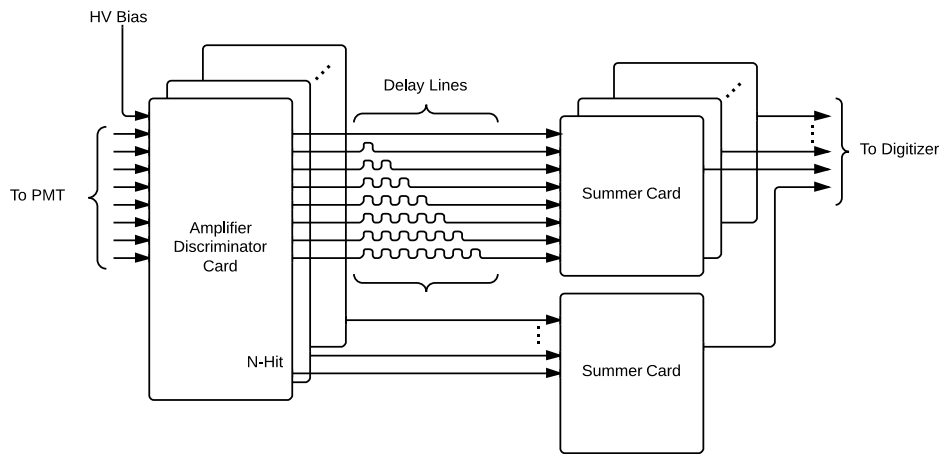}
\caption{Block Diagram of MiniCLEAN Veto Electronics
\label{fig:block_diagram}}
\end{center}
\end{figure}

The electronic components of the veto subsystem are the PMT high voltage
supply, the PMT signal electronics, and the data acquisition
electronics. A schematic showing the system outline is shown in
Figure~\ref{fig:block_diagram}.

The 48 veto PMTs use 6 channels of a 12-channel card of the
MiniCLEAN inner liquid argon detector PMT high voltage supply (LeCroy 1461 VISyN)
with SHV connectors. The voltage divider circuits to distribute power
from each of 6 HV channels to 8 PMTs are part of the amplifier
discriminator card described in Section~\ref{sec:Amp-Disc}. The veto
PMT signal electronics consist of 6 amplifier discriminator boards which each service
8 PMTs and 7 summer boards designed and built by MIT Bates Research and Engineering Center.
The veto PMTs use 6 channels on one CAEN V1720 digitizer
card of the MiniCLEAN DAQ system. The veto signal multiplexing
electronics are part of the summer card described
Section~\ref{sec:Sum}. The CAEN DAQ is described further
in~\cite{ref:gastler_thesis}.

\subsection{Amplifier discriminator board}
\label{sec:Amp-Disc}
\begin{figure}[ht]
\begin{center}
\includegraphics[width=5.5in, keepaspectratio=true, page=2, trim=4.54in 2.12in 4.54in 2.12in, clip=true]{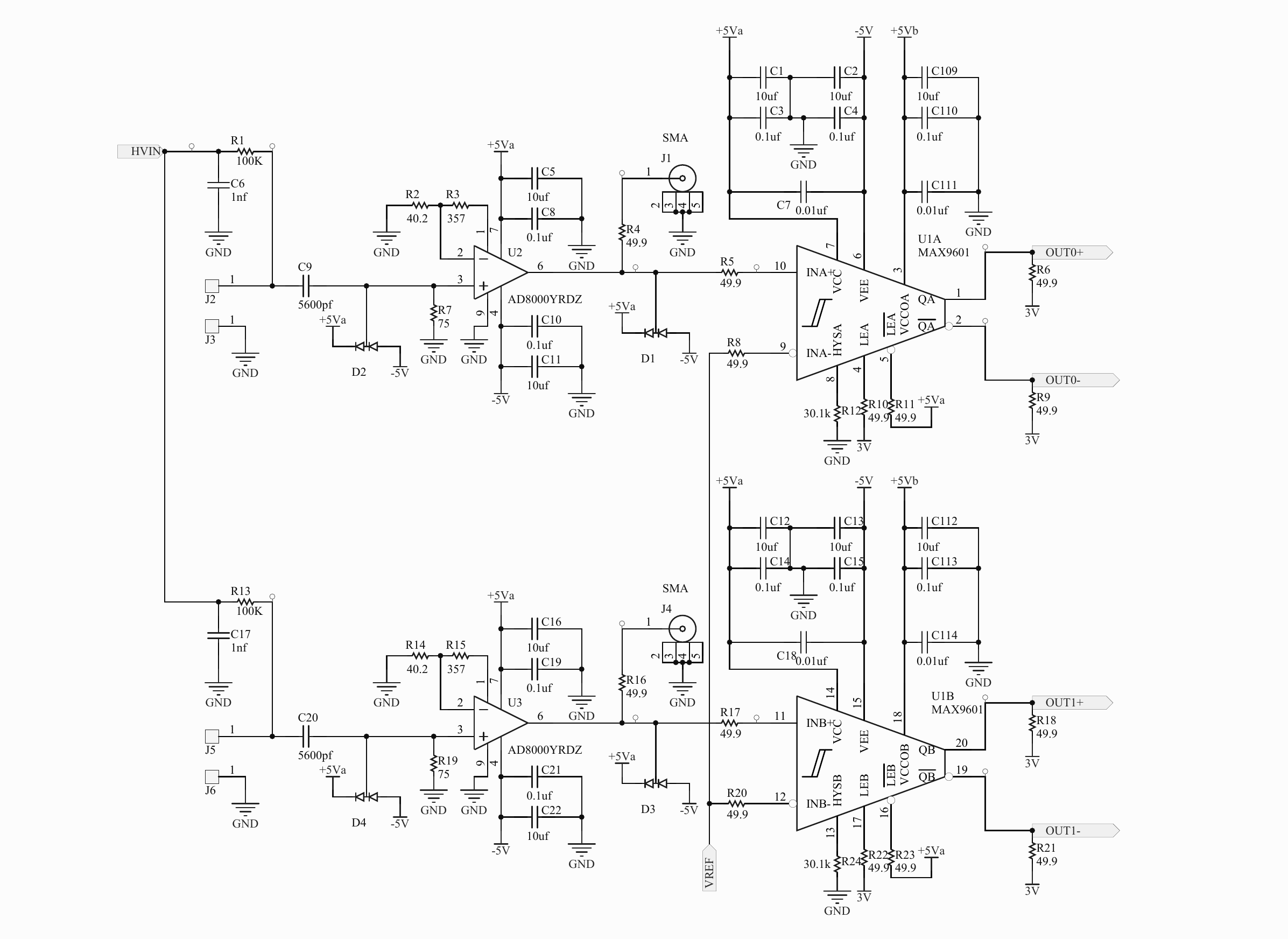}
\caption{Two channels of the amplifier discriminator card. J2 and J5 make the signal connections to the PMTs. J1 and J4 make the connections to the delay lines. U2 and U3 are the amplifiers, U1 is the two channel comparator.
\label{fig:ampdiscsch}}
\end{center}
\end{figure}

The amplifier discriminator board is the front end electronics board.
The board provides high voltage distribution to the PMTs, signal
amplification, and signal discrimination for the N-Hit sum. The PMT
bias voltage is fed to each of the PMT connections through a single
SHV connector. Each PMT capacitively couples the signal onto the bias
voltage line. On the amplifier discriminator board each channel has
an amplifier input capacitively coupled to a PMT bias line. The schematic for the
PMT bias scheme, along with two channels of the amplifier discriminator, is shown in Figure~\ref{fig:ampdiscsch}.

The HV bias components were surface mount types. Because of the
aspect ratio of the width and length of some of the components, the
components had to be raised slightly off of the board during the
soldering process to ensure that flux could be cleaned from underneath
the components. In addition, the HV components were potted with
corona dope after assembly.

A high bandwidth current feedback amplifier (CFA) provides the large
gain and bandwidth needed to amplify the pulses by a factor of 10 in a
single stage. We selected the Analog Devices AD8000 CFA to amplify
the PMT pulses. The amplified pulses are fed out to the delay
lines and to the on-board comparators. Due to the high gain
requirements, and the variations between the CFAs, all of the parts
installed were tested beforehand and binned into parts with a similar
DC offset. During board assembly, care was taken to ensure that all
of the amplifiers installed on a single board had a DC offset which varied by no more than 10\%.

The comparator compares each channel with a programmable threshold
voltage, common to all 8 channels on each board. The output of the
comparator is a fast differential PECL output. These are summed with
a differential amplifier and output using a transformer as a single
ended signal, the "N-Hit sum" for those 8 channels.

The amplifier discriminator board has local linear power regulation.
Each board has two +5~V supplies, one -5~V supply, and one +3~V supply.
The CFA runs on the +5~V and -5~V supplies. The comparator runs on the
second +5~V supply and the +3~V supply (necessary for PECL output).
The two +5~V supplies isolate the output supply for
the comparator in order to reduce cross talk. The regulators are fed
by a common bipolar supply and ground nominally at $\pm$9~V. The
nominal values for current draw are 500~mA on the positive rail and
250~mA on the negative rail.

The board is constructed with 4 copper layers and standard FR4
substrate material, with 0.062 inch thickness. It is mounted on a piece
of FR4 to protect HV nets on the bottom of the board from arcing and
to fit it into a 6U VME style crate. 9 SHV connectors are mounted on
a front panel connected to the FR4 and provide the HV bias input and
eight PMT connections. These PMT connections attach on the left side
of the board with short insulated wire pig tails. On the right
side of the board 9 SMA connectors are board mounted. The SMA
connectors provide the 8 signal outputs to the delay lines and the
N-Hit sum signal. A photo of the amplifier discriminator
board is shown in Figure~\ref{fig:boards}.

\subsection{Delay lines}
\label{sec:Delay}
\begin{figure}[ht]
\begin{center}
	\includegraphics[height=2in, keepaspectratio=true]{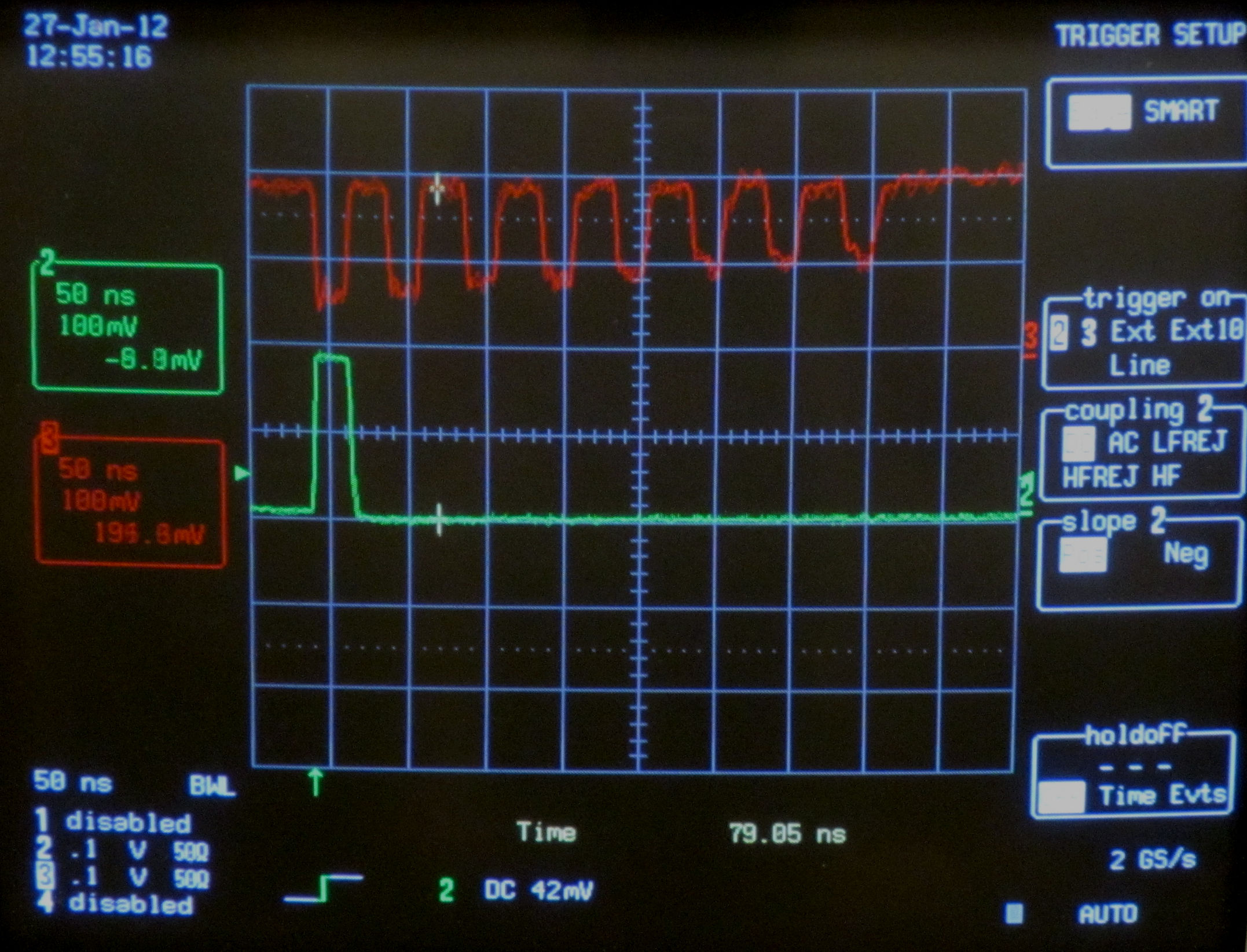}
	\caption{Oscillicope capture from the output of a summer board. Eight channels
		pulsed with a 10~mV synchronous pulse into the amplifier discriminator board. The amplified
		signals are routed through the delay lines into the summer board. The top trace is
		the output of the summer board. The bottom trace is the N-Hit sum output from the
		amplifier discriminator board, used as a trigger.
\label{fig:multipulse}}
\end{center}
\end{figure}

The delay lines connect from the amplifier discriminator boards to the
summer boards. They are designed to delay the signals between 50~ns and 350~ns.
Since the signal pulses are expected to be short in
comparison to the 50~ns delay window, delaying the signals and summing
them together has the effect of time multiplexing the signals. The
delay lines are made of different lengths of Belden 9310 cable fitted
with male SMA connectors. The cable was chosen for its low
attenuation, slow signal propagation speed, and small bend radius.
There are 7 different lengths ranging form 32~ft to 224~ft, with every 32~ ft
equating to about 50~ns of delay. These are coiled inside a metal box
with pig tails to connect to the amplifier discriminator boards and
the summer boards. The metal box satisfies fire protection concerns
in the mine, as the Belden 9310 is not plenum rated. There are two
sets of each of the seven lengths in each box. One of the signals has
a $\sim$0~ns delay and is connected from the amplifier discriminator card to
the summer card with a very short jumper cable. Figure~\ref{fig:multipulse}
shows the time multiplexed ouput of 8 channels fed with a synchronous pulse.

\subsection{Summer board}
\label{sec:Sum}
\begin{figure}[ht]
	\begin{center}
		\includegraphics[height=2in, keepaspectratio=true]{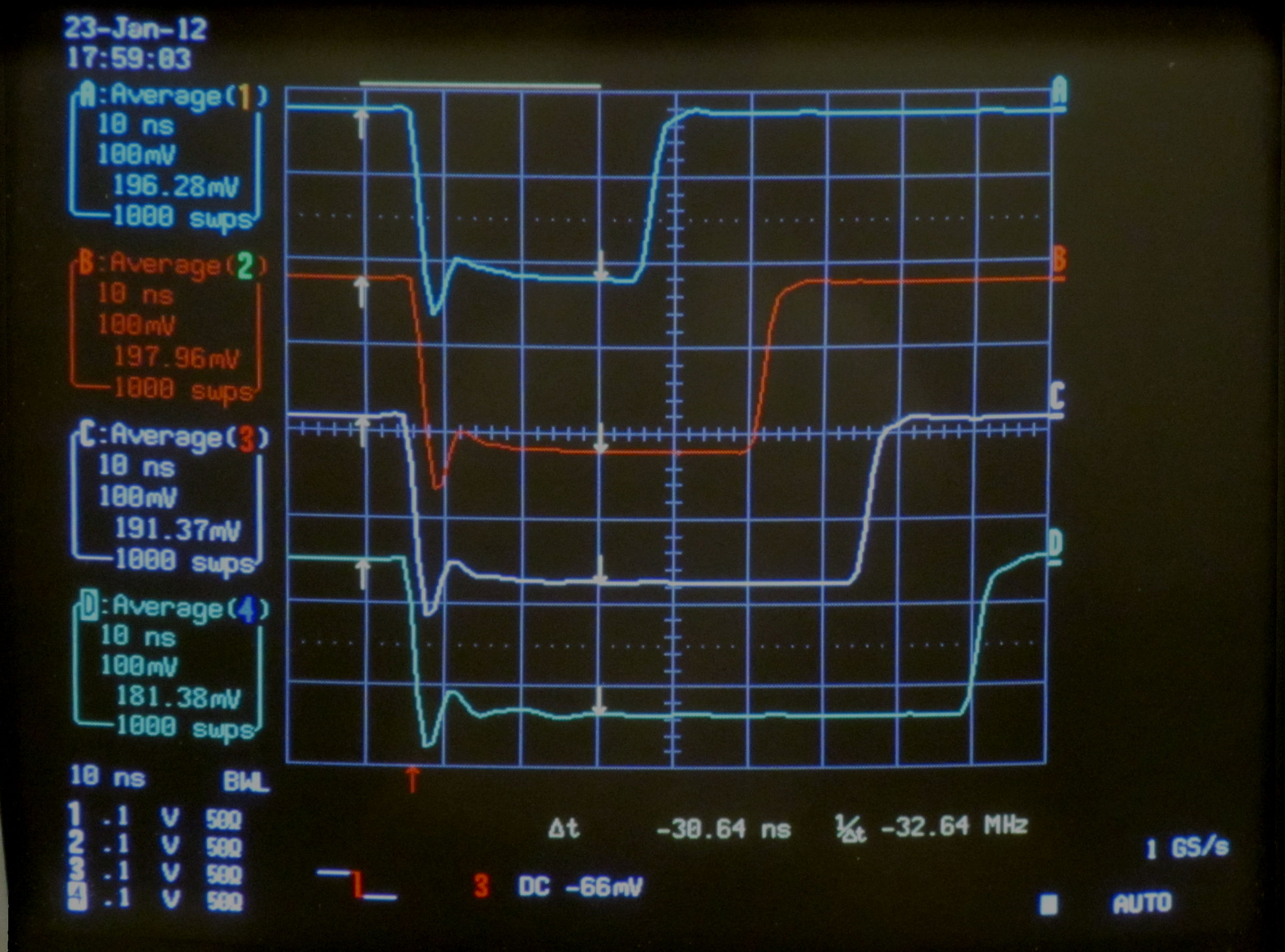}
		\includegraphics[height=2in, keepaspectratio=true]{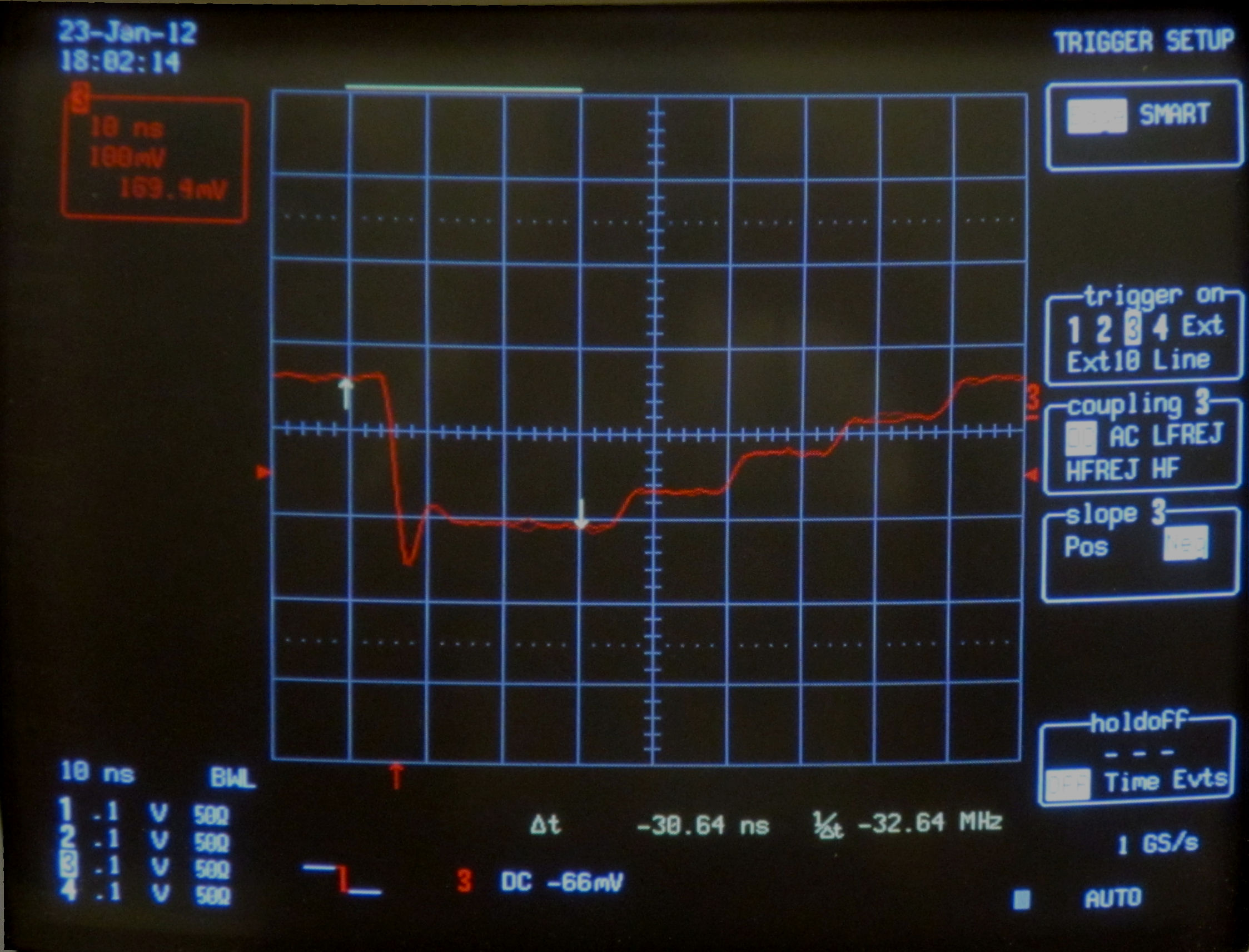}
		\caption{Input (left) and output (right) to the summer board measured with a test pulse generator.
		\label{fig:summerpulse}}
	\end{center}
\end{figure}

\begin{figure}[ht]
\begin{center}
\includegraphics[width=4in, keepaspectratio=true]{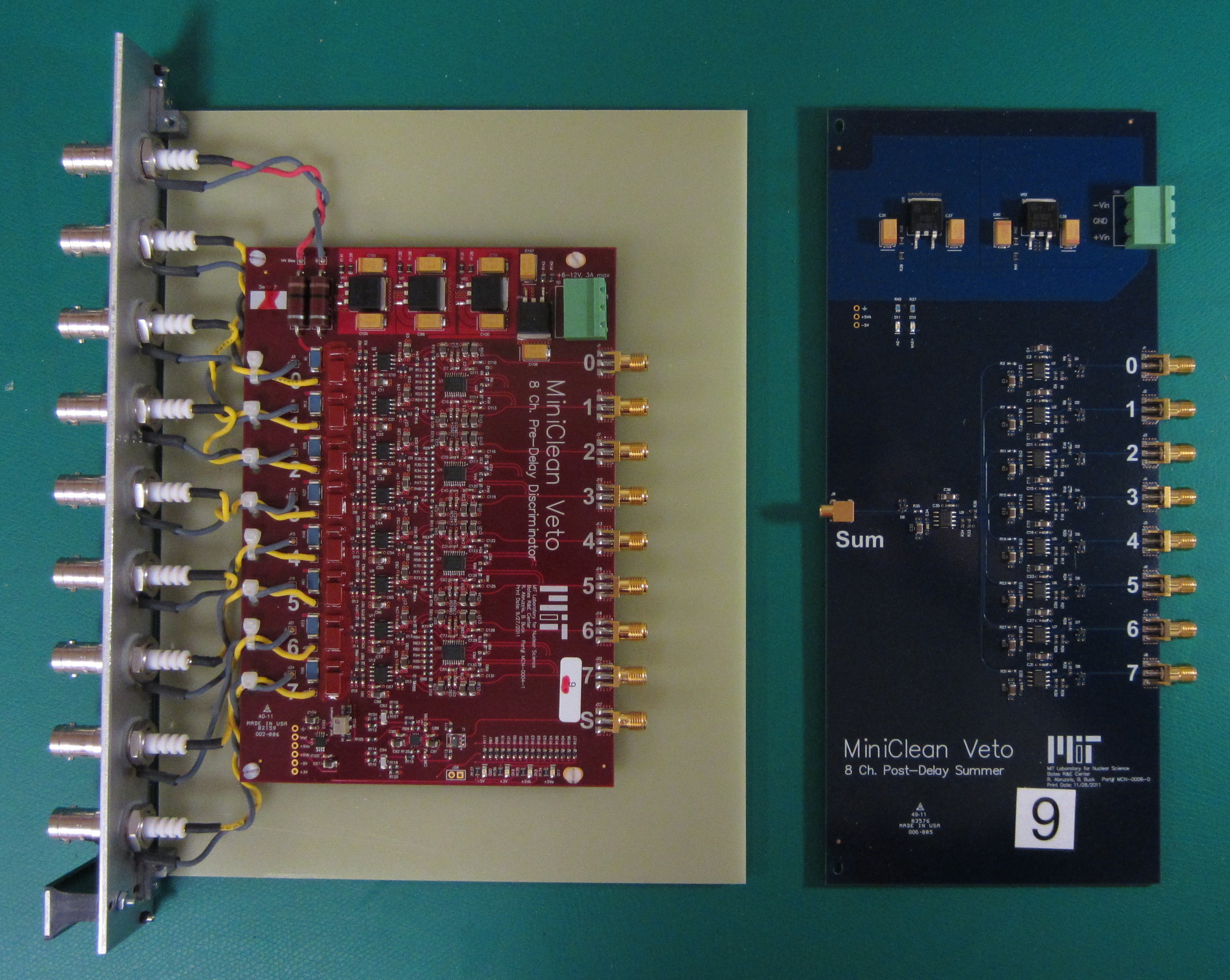}
\caption{Amplifier discriminator board (left) and summer board (right).
\label{fig:boards}}
\end{center}
\end{figure}

The summer board sums the analog channels after they have been
delayed and also sums the N-Hit sums from each of the 6 cards to make
a global N-Hit sum. The summer board has 8 inputs which are female
SMA connectors. Each of these inputs is fed, DC coupled, into a CFA
to provide buffering. This also provides the opportunity to fine tune
the gain and equalize the signals after the cable delay by adjusting
the two feedback resistors. the outputs are fed into another CFA
configured as an analog summing node. This outputs through a series
resistor to an female MCX connector. A short jumper cable connects
the summer board output to the digitizer input. Like the amplifier
discriminator board, the summer board has local power regulation.
Each board has one +5~V supply and one -5~V supply. The regulators are
fed by a common bipolar supply nominally at $\pm$9~V.
The nominal values for current draw are 250~mA on the positive
rail and 250~mA on the negative rail. A photo of the summer
board is shown in Figure~\ref{fig:boards}, an example test pulse run
through the summer board is shown in Figure~\ref{fig:summerpulse}.

\subsection{Housing and assembly}
\label{sec:Housing}
The electronics are housed in a rack which holds a crate for the
electronics, a fan unit for cooling, a power supply unit, and all of
the delay lines. The power supply module provides $\pm$9~V up to
130~W. Estimated total power draw is 72~W. The power supply uses two
discrete 9~V switching power supplies which provide +9~V, 0~V, and -9~V
power rails running on the back of the rack. Connectors are attached
to these rails and insert into each of the electronics boards. Each
of the components was tested individually at MIT. Amplifier
discriminator boards were high-pot tested and all channels were tested
to ensure uniformity. Summer boards were similarly tested. The
boards were assembled into a VME style crate and mounted in a rack
together with the power supply unit, a fan unit, and the delay
lines. An integration test was done with the MiniCLEAN electronics
at Boston University to
demonstrate the system running while attached to the CAEN DAQ system.

\section{Summary}
\label{Summary}
A full test of the electronics chain using simulated signals,
integrated with the MiniCLEAN DAQ, demonstrated the veto
system working properly. We were able to see simulated events pushed
into the system and see the N-Hit trigger signal increase
proportionally to the number of PMTs which were hit. We could observe
the output signals on the time multiplexed outputs from the summer
boards (Figure~\ref{fig:multipulse}).
Once operational, the full system will
multiplex 48 PMT signals into 6 digitizer channels and provide a
system wide N-Hit sum. The system wide N-Hit will provide the veto
trigger to the MiniCLEAN DAQ. The system is now at SNOLAB ready for
implementation in the experiment.

\acknowledgments
The authors would like to acknowlege support from NSF grant PHY-0970047 and the MIT Bates Research and Engineering Center.

\end{document}